\documentclass{article}

\usepackage{balance}  % to better equalize the last page
\usepackage{graphics} % for EPS, load graphicx instead
\usepackage{graphicx}
\usepackage{epstopdf}
\usepackage{times}    % comment if you want LaTeX's default font
\usepackage{url}      % llt: nicely formatted URLs
\usepackage{algorithmic}
\usepackage{algorithm2e}
\usepackage{amsmath}
\usepackage{subfig}
\usepackage{authblk}

\begin{document}

\title{Distributed Matrix Factorization using Asynchrounous Communication}

\author{Tom Vander Aa}
\author{Imen Chakroun}

\affil{
    Exascience Lab,
    Imec,
    Kapeldreef 75,
    B-3001 Leuven,
    Belgium \\
    \{tom.vanderaa,imen.chakroun\}@imec.be
}

\author{Tom Haber}

\affil{
  Expertise Centre for Digital Media,
  Wetenschapspark 2,
  3590 Diepenbeek,
  Belgium \\
  {tom.haber@uhasselt.be}
}
                      
\maketitle

\begin{abstract}
Using the matrix factorization technique in machine learning is very common
mainly in areas like recommender systems. Despite its high prediction accuracy
and its ability to avoid over-fitting of the data, the Bayesian Probabilistic
Matrix Factorization algorithm (BPMF) has not been widely used on large scale
data because of the prohibitive cost. In this paper, we propose a distributed
high-performance parallel implementation of the BPMF using Gibbs sampling on
shared and distributed architectures. We show by using efficient load balancing
using work stealing on a single node, and by using asynchronous communication in
the distributed version we beat state of the art implementations.
\end{abstract}

\section{Introduction}
\label{introduction}

Recommender Systems (RS) have become very common in recent years and are useful
in various real-life applications.
% reference

The most popular ones are probably suggestions for movies on Netflix and books
for Amazon. However, they can also be used in more unlikely area such drug
discovery where a key problem is the identification of candidate molecules that
affect proteins associated with diseases. One of the approaches that have been
widely used for the design of recommender systems is collaborative filtering
(CF). This approach analyses a large amount of information on some users'
preferences and tries to predict what other users may like. A key advantage of
using collaborative filtering for the recommendation systems is its capability
of accurately recommending complex items (movies, books, music, etc) without
having to understand their meaning.  For the rest of the paper, we refer to the
items of a recommender system by movie and user though they may refer to
different actors (compound and protein target for the ChEMBL benchmark for
example \cite{ChEMBL}).

% Properties of the matrices

\begin{figure}
\begin{minipage}{.5\textwidth}
    \centering
    \includegraphics[width=\columnwidth]{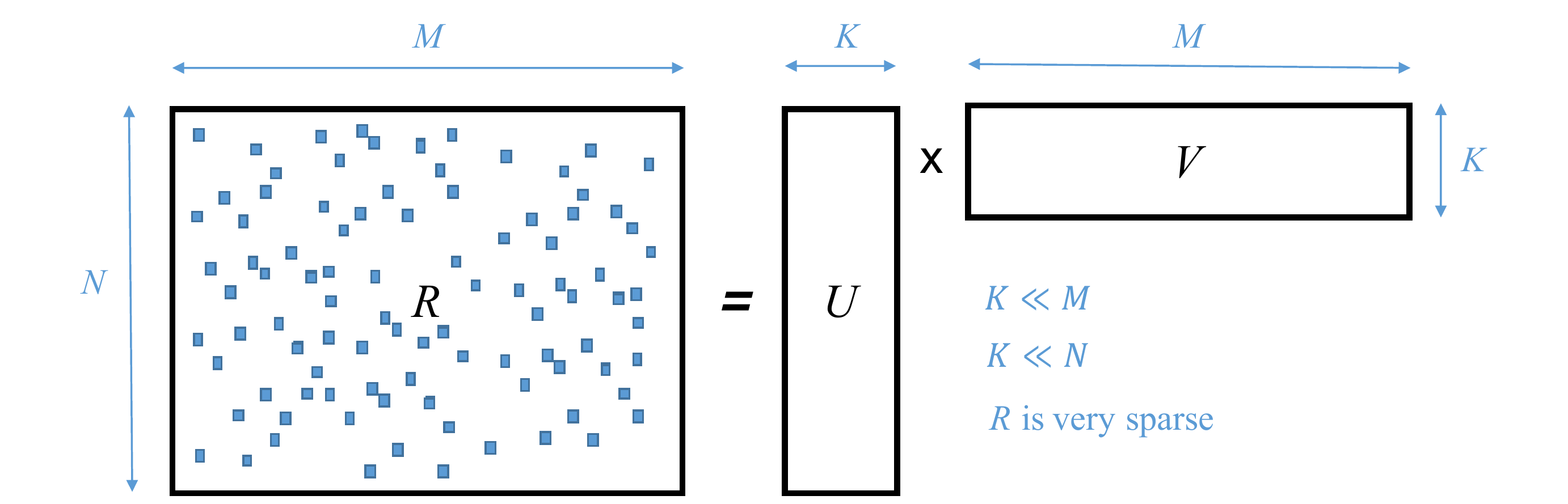}
    \caption{Low-rank Matrix Factorization}
    \label{fig:mf}
\end{minipage}
\begin{minipage}{.5\textwidth}
    \centering
    \includegraphics[width=\columnwidth]{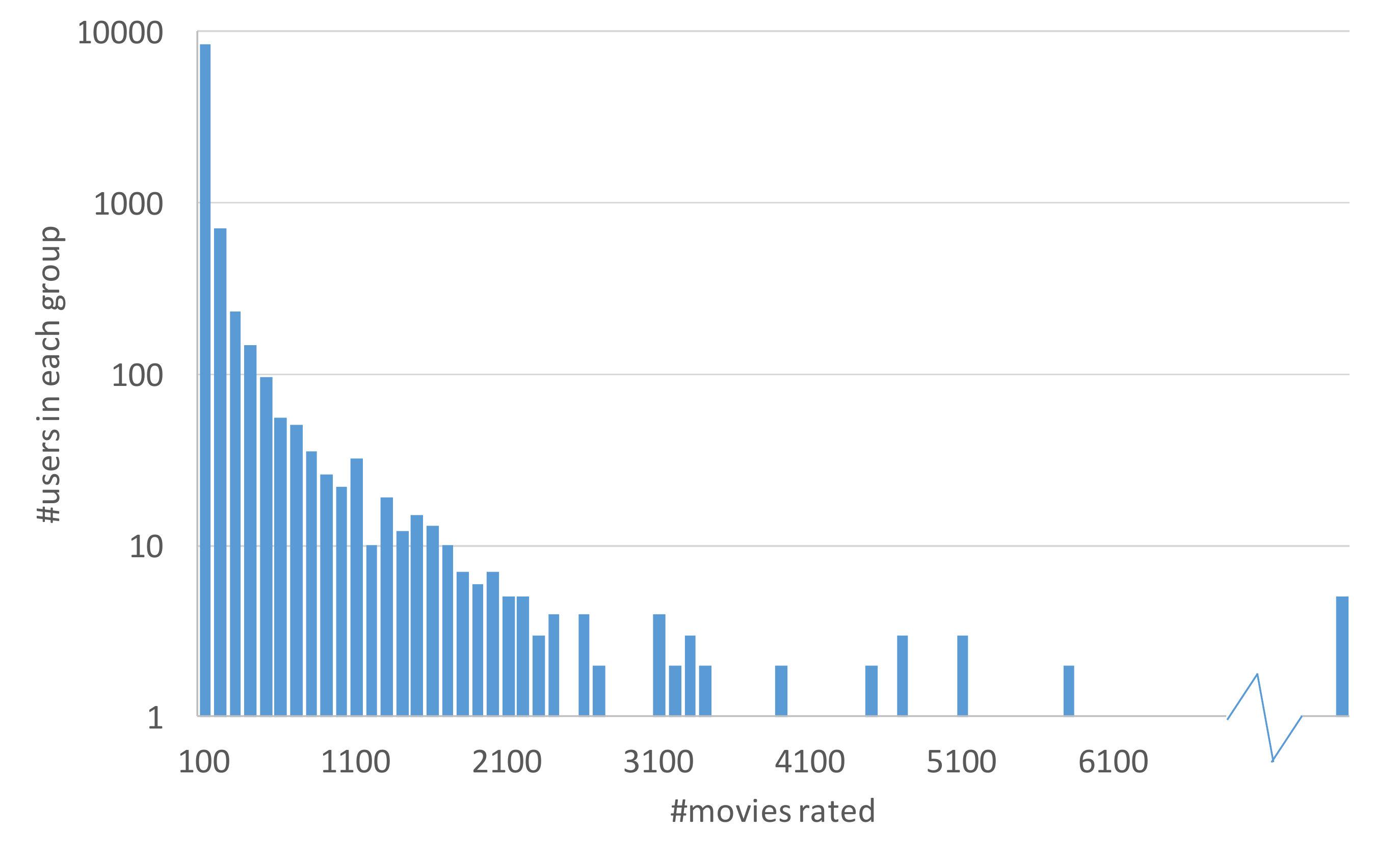}
    \caption{Histogram for the ChEMBL dataset of the number of ratings per user.}
    \label{fig:chembl_histogram}
\end{minipage}
\end{figure}

To deal with collaborative filtering challenges such as the size and the
sparseness of the data to analyze, Matrix Factorization (MF) techniques have
been successfully used. Indeed, they are usually more effective because they
take into consideration the factors underlying the interactions between users
and movies called \emph{latent features}. As sketched in Figure~\ref{fig:mf},
the idea of these methods is to approximate the user-movie rating matrix $R$ as
a product of two low-rank matrices $U$ and $V$ (for the rest of the paper $U$
refers to the users matrix and $V$ to the movie matrix) such that $R \approx
U \times V$. In this way $U$ and $V$ are constructed from the known ratings in
$R$, which is usually very sparsely filled. The recommendations can be made from
the approximation $U \times V$ which is dense.  If $M$ $\times$ $N$ is the
dimension of $R$ then $U$ and $V$ will have dimensions $M$ $\times$ $K$ and $N$
$\times$ $K$.  $K$ represents then number of latent features characterizing the
factors, $K \ll M$, $K \ll N$.

Popular algorithms for low-rank matrix factorization are alternating
least-squares (ALS) \cite{parallelALS}, stochastic gradient descent (SGD)
\cite{parallelSGD} and the Bayesian probabilistic matrix factorization (BPMF)
\cite{BPMF}. Thanks to the Bayesian approach, BPMF has been proven to be
more robust to data-overfitting and released from cross-validation (needed for
the tuning of regularization parameters). In addition, BPMF easily incorporates
confidence intervals and side-information \cite{SIDEINFORMATION, simm:macau}.
Yet BPMF is more computational intensive and thus more challenging to implement
for large datasets.  Therefore, the contribution of this work is to propose a
parallel implementation of BPMF that is suitable for large-scale distributed
systems. An earlier version of this work has been published at
\cite{bpmf_cluster16}. Compared to that earlier version, this works adds
efficient asynchronous communication using GASPI~\cite{grunewald:gaspi}.

The remainder of this paper is organized as follows.  Section~\ref{sec:BPMF}
describes the BPMF algorithm. In Section~\ref{sec:MBPMF}, the shared-memory
version of the parallel BPMF is described. In Section~\ref{sec:DBPMF}, details
about the distributed BPMF are given. The experimental validation and associated
results is presented in Section~\ref{sec:experiments}.  In
Section~\ref{sec:related} existing work dealing with parallel matrix
factorization techniques and BPMF in particular is presented. Some conclusions
and perspectives of this work are drawn in Section~\ref{sec:conclusion}

\section{BPMF} \label{sec:BPMF}

% how the algo works

The BPMF algorithm \cite{BPMF} puts matrix factorization in a Bayesian
framework by assuming a generative probabilistic model for ratings with prior
distributions over parameters.  It introduces common multivariate Gaussian
priors for each user of $U$ and movie in $V$. To infer these two priors from the
data, BPMF places fixed uninformative Normal-Wishart hyperpriors on them. We use
a Gibbs sampler to sample from the prior and hyperprior distributions.

This sampling algorithm can be expressed as the pseudo code shown in
Algorithm~\ref{algo:bpmf_pseudo}. Most time is spent in the loops updating $U$
and $V$, where each iteration consist of some relatively basic matrix and
vector operations on $K \times K$ matrices, and one computationally more
expensive $K \times K$ matrix inversion.

\begin{algorithm}
    \footnotesize
  \LinesNumbered
  \For{sampling iterations}
  {
      sample hyper-parameters movies based on V

      \For{all movies $m$ of $M$}
      {
          update movie model $m$ based on
          ratings ($R$) for this movie and 
          model of users that rated this movie,
          plus randomly sampled noise
      }

      sample hyper-parameters users based on U

      \For{all users $u$ of $U$}
      {
          update user $u$ based on
          ratings ($R$) for this user and 
          model of movies this user rated,
          plus randomly sampled noise
      }

      \For{all test points}
      {
          predict rating and compute RMSE
      }

  }
  \caption{BPMF Pseudo Code} 
  \label{algo:bpmf_pseudo}
\end{algorithm}% 

These matrix and vector operations are very well supported in
Eigen~\cite{Eigen} a high-performance modern C++11 linear algebra library.
Sampling from the basic distributions is available in the C++ standard template
library (STL), or can be trivially implemented on top. As a results the
Eigen-based C++ version of Algorithm~\ref{algo:bpmf_pseudo} is a mere 35 lines
of C++ code. 

\section{Multi-core BPMF}
\label{sec:MBPMF}

In this section we describe how to optimize this implementation to run
efficiently on a shared memory multi-core system. A version for distributed
systems with multiple compute nodes is explained in a separate section.

\subsection{Single Core Optimizations}

Most of time is spent updating users' and movies' models. This involves
computing a $K \times K$ outer product for the covariance matrix and inverting this
matrix to obtain the precision matrix. Since the precision matrix is used only
once, in a matrix-vector product, we can avoid the full inverse and only compute
the Cholesky decomposition. Furthermore, if the number of ratings for a
user/movie is small a rank-one update~\cite{stewart:matrix_algos} is more
efficient. 

Updating a single user in $U$ depends on the movies in $V$ for whom there are
ratings in $R$, Hence, the access patterns to $U$ and $V$ are determined by the
sparsity pattern in $R$. By reordering the columns and rows of $R$, we can
improve the data locality and thus the program's cache behavior. Since the
access pattern in BPMF is similar to access pattern in a Sparse Matrix-Vector
Multiplication (SPMV), we reused the technique proposed in
\cite{vastenhouw:mondriaan}. 

To sample from the hyper-parameters a global average and covariance across both
$U$ and $V$ needs to be computed. Standalone, the computation of these values is
dominated by the long-latency memory accesses to $U$ and $V$. However, if we
integrate the computation of these aggregates with the updates of $U$ and $V$,
they become almost free.

\subsection{Multi-core-based parallel BPMF}

% shared memory parallelism: shared matrices
The main challenges for performing BPMF in parallel is how to distribute the
data and the computations amongst parallel workers (threads and/or distributed
nodes).  For the shared memory architectures, our main concerns where using as
many threads as possible, keeping all threads as busy as possible and minimizing
memory discontinuous accesses.  Since the number of users entries (resp. movie
entries) are very large and since they can all be computed in parallel, it make
sense to assigned a set of items to each thread. 

\begin{figure}
    \centering
\includegraphics[width=0.6\textwidth]{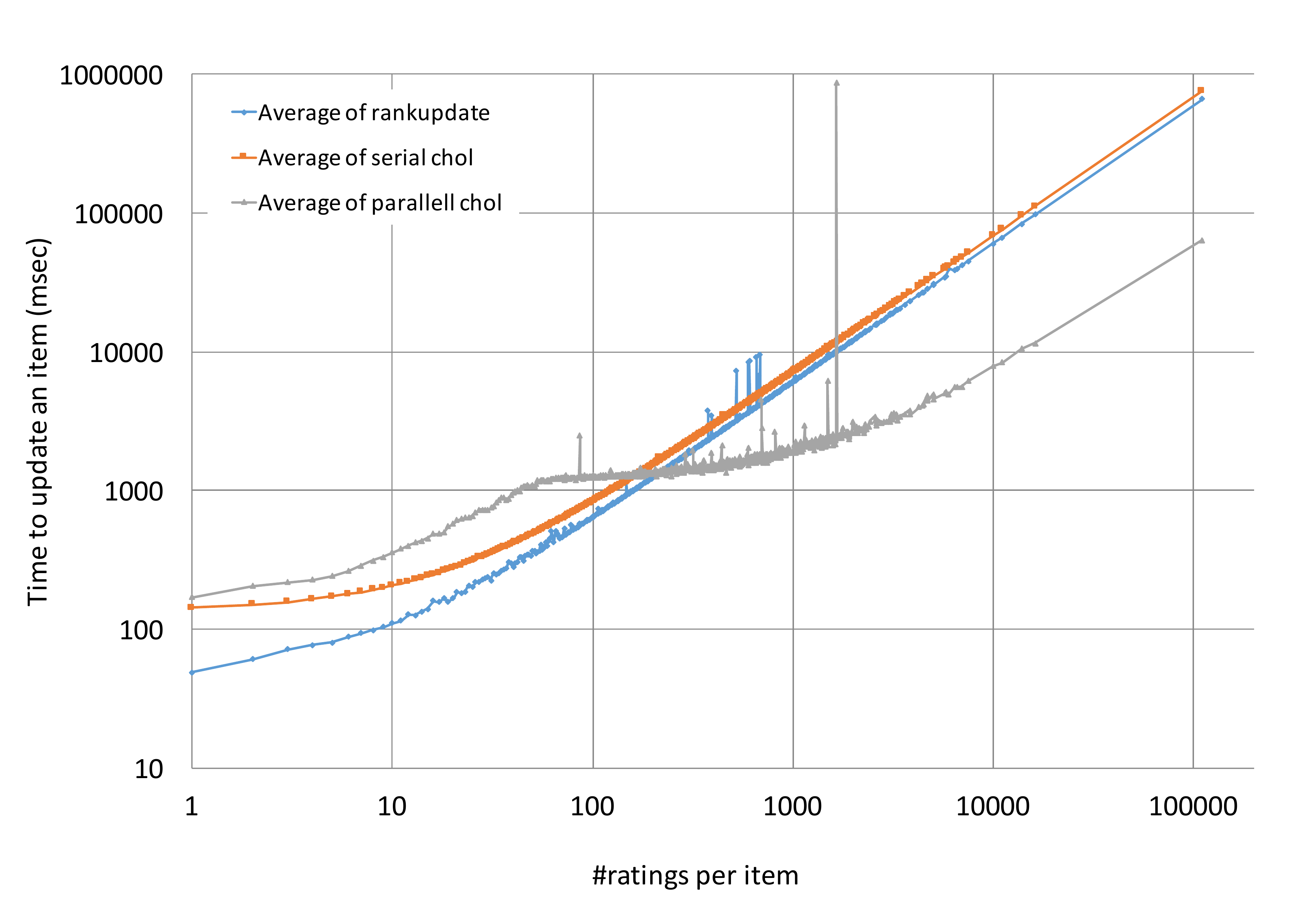}
\caption{Compute time to update one item for the three methods: sequential
rank-one update, sequential Cholesky decomposition, and parallel Cholesky
decomposition as a function of the number of ratings.}
\label{fig:parallel_sample}
\end{figure}

Next, balanced work sharing is a major way of avoiding idle parallel threads.
Indeed, if the amount of computations is not balanced some threads are likely to
finish their tasks and stay idle waiting for others to finish.  As can be seen
in Figure~\ref{fig:chembl_histogram}, there are items (users or movies) with a large
number of ratings and for whom the amount of compute is substantially larger
than those items with less ratings. To ensure a good load balance, we use a
cheaper but serial algorithm using the aforementioned rank-one update, for items
with less than 1000 ratings. For items with more ratings, we use a parallel
algorithm containing a full Cholesky decomposition. This choice is motivate by
Figure~\ref{fig:parallel_sample} which shows the time to update one item versus
the number of ratings for the three possible algorithms.  By using the parallel
algorithm for more expensive users/movies we effectively split them up in more
smaller tasks that can utilize multiple cores on the system.

\section{Distributed parallel BPMF} \label{sec:DBPMF}

The multi-core BPMF implementation presented above has been extended to
distributed systems using three different distributed programming models:
MPI~\cite{MPI}, GASPI~\cite{grunewald:gaspi} and ExaSHARK~\cite{SHARK}. In this section we
first describe the three programming models, next how the data is distributed
across nodes, how the work per node is balanced and how communication is
handled, for the three approaches.

\subsection{Distributed Programming}

\subsubsection{MPI-3.0}
Message Passing Interface (MPI) is a standardized and portable message-passing
system for distributed systems. The latest standard MPI-3.0 includes
features important for this BPMF implementation, for example: support for
asynchronous communication, support for hybrid application combining message
passing with shared memory level parallelism like OpenMP~\cite{openmp} or TBB~\cite{tbb}.

\subsubsection{GASPI}
The Global Address Space Programming Interface (GASPI \cite{grunewald:gaspi}) is the
specification for a PGAS style programming model for C/C++ and Fortran. The API
consists of a set of basic routines. As an alternative to MPI, its main
advantages are \emph{i)} its one-sided communication layer that can take full
advantage of the hardware capabilities to utilize remote direct memory access
(RDMA) for spending no CPU cycles on communication, \emph{ii)} the fact that the
GASPI library has been optimized to work in a multi-threaded environment and
\emph{iii)} its seamless interoperability with MPI.

% ExaSHARK
\subsubsection{ExaSHARK}
Compared to MPI and GASPI, ExaSHARK is a much higher abstraction level library
designed to handle matrices that are physically distributed across multiple
nodes. The access to the global array is performed through logical indexing.
ExaSHARK is portable since it is built upon widely used technologies such as MPI
and C++ as a programming language. It provides coding via a global-arrays-like
interface which offers template-based functions (dot products, matrix
multiplications, unary expressions) which offers transparent execution across
the whole system. 

\subsection{Data Distribution}

We distribute the matrices $U$ and $V$ across the system where each nodes
computes their part.  When an item is computed, the rating matrix $R$ determines
to what nodes this item needs to be sent.

Our main optimization concern on how to distribute $U$ and $V$ is to make sure
the computational load is distributed equally as possible and the amount of data
communication is minimized.  Similarly to the cache optimization mentioned
above, we can reorder the rows and columns in $R$ to minimize the number of
items that have to be exchanged, if we split and distribute $U$ and $V$
according to consecutive regions in $R$.

Additionally we take work balance in to account when reordering $R$. For this we
use a workload model derived from Figure~\ref{fig:parallel_sample}. The blue
curve in the figure give a reasonable idea of the amount of work for a user or
movie in relation to the amount of ratings. As you can see, when the number of
ratings is small, the work per rating is higher than for items with many
ratings. Hence we approximate the workload per user/movie with fixed cost,
plus a cost per movie rating.

\subsection{Updates and data communication}

\subsubsection{Communication using ExaSHARK} For the users updates, only one-sided
communication is used in the case a user is outside a process range, namely the
\texttt{\small GlobalArray::get()} routine. Indeed, thanks to the PGAS model, each
process knows which other process owns a particular range of the global array. 

\subsubsection{Communication using pure MPI}
To allow for communication and computation to overlap we send the updated
user/movie as soon as it has been computed. For this we use the asynchronous MPI
3.0 routines \texttt{\small MPI\_Isend} and \texttt{\small MPI\_Irecv}. However, the overhead
of calling these routines is too much to individually send each item to the
nodes that need it. Additionally, too many messages would be in flight at the
same time for the runtime to handle this efficiently. Hence we store items that
need to be sent in a temporary buffer and only send when the buffer is full.
% Figure~\ref{fig:mpi_buffered} below show the general scheme.
% 
% \begin{figure}
% \includegraphics[width=\columnwidth]{mpi_buffered}
% \caption{Buffering items before they are sent}
% \label{fig:mpi_buffered}
% \end{figure}

\subsubsection{Communication using GASPI}
Because GASPI is more light-weight, we can afford to simply send (gaspi\_write)
an item once it has been computed. 

\section{Validation} \label{sec:experiments}

In this section, we present the experimental results and related discussion for
the proposed parallel implementations of the BPMF described above. 

\subsection{Hardware platform} 

We performed experiments on Lynx a cluster with 20 nodes, each equipped with dual
6-core Intel(R) Westmere CPUs with 12 hardware threads each, a clock speed
2.80GHz and 96 GB of RAM, and on Anselm a cluster with 209 nodes, each node equipped
with 2 8-core Intel(R) Sandy Bridge CPUs with at least 64GB RAM per node.

%and on Fermi, an IBM BlueGene/Q system with 10240 nodes, each equipped with 16
%cores running at 1.2Ghz and 16 GB of memory.

\subsection{Benchmarks} \label{bench}

Two public benchmarks have been used to evaluate the performances of the
proposed approaches: the ChEMBL dataset \cite{ChEMBL} and the
MovieLens~\cite{harper:movielens} database. 

The ChEMBL dataset is related to the drug discovery research field. It contains
descriptions for biological activities involving over a million chemical
entities, extracted primarily from scientific literature. Several version exist
since the dataset is updated on a fairly frequent basis. In this work, we used
a subset of the version 20 of the database which was released on February 2015.
The subset is selected based on the half maximal inhibitory concentration
(IC50) which is a measure of the effectiveness of a substance in inhibiting a
specific biological or biochemical function. The total ratings number is around
1023952 from 483500 compounds (acting as users) and 5775 targets (acting as
movies). 

The MovieLens dataset (ml-20m) describes 5-star rating and free-text tagging
activity from MovieLens, a movie recommendation service.  It contains 20M
ratings across 27278 movies. These data were created by 138493 users between
January 09, 1995 and March 31, 2015.

For all the experiments, all the versions of the parallel BPMF reach the same
level of prediction accuracy evaluated using the root mean square error metric
(RMSE) which is a used measure of the differences between values predicted by a
model or an estimator and the values actually observed \cite{Hyndman:RMSE}.

\subsection{Results for Multi-core BPMF}

In this section, we compare the performance of the proposed multi-core
BPMF with the Graphlab library which is a state of the art library widely used
in machine learning community. We have chosen GraphLab because it is known to
outperform other similar graph processing implementations~\cite{Guo:graphlab}.

\begin{figure}
    \centering
\includegraphics[width=0.6\columnwidth]{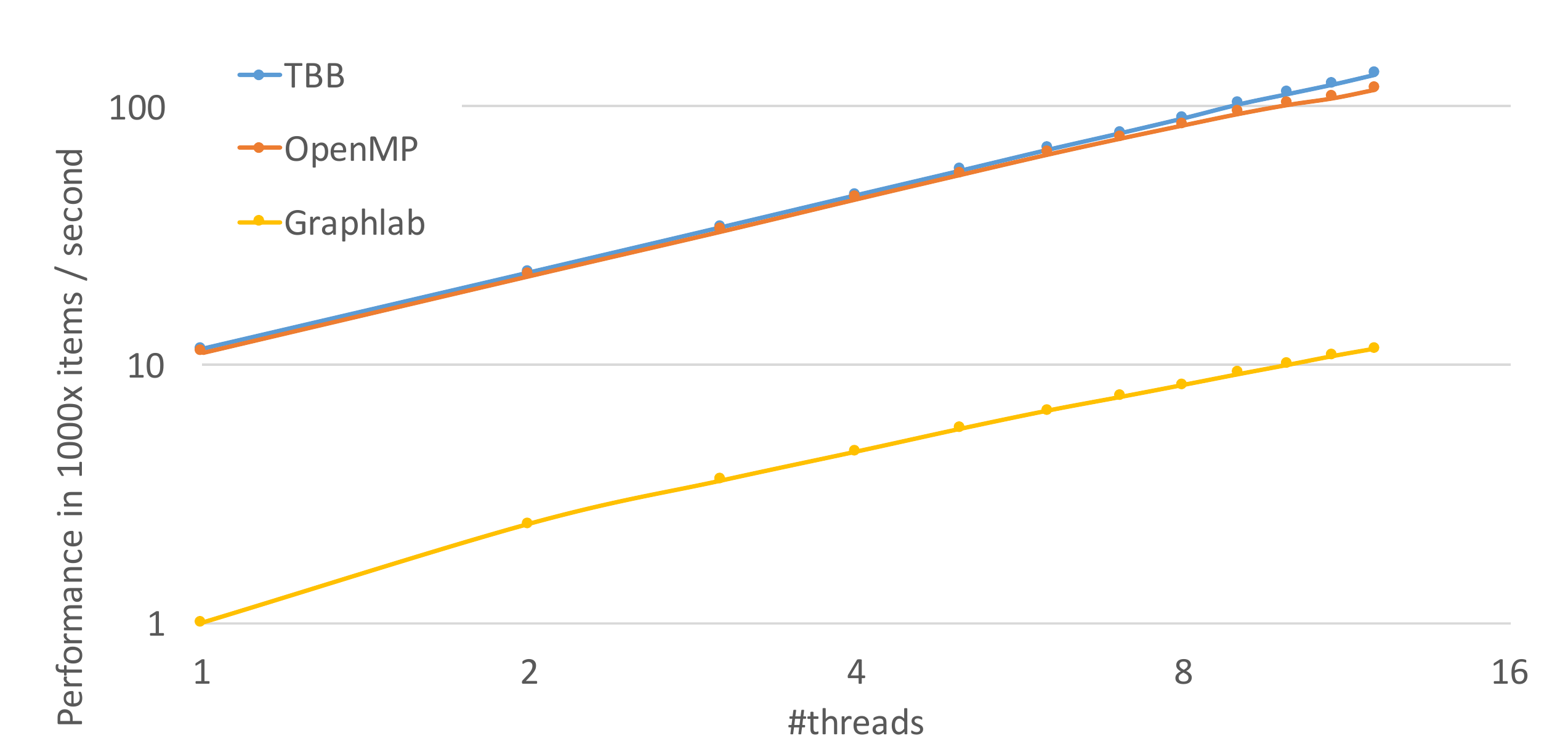}
\caption{Performance of the multi-core BPMF on the ChEMBL dataset in number of
updates to $U$ and $V$ versus the number of parallel threads.}
\label{fig:multicore}
\end{figure}

The results presented in Figure \ref{fig:multicore} report the performance in
number of updates to $U$ and $V$ per second for the ChEMBL benchmark suite on a
machine with 12 cores for three different version: \textbf{TBB} The C++
implementation using Intel's Threading Building Blocks (TBB) for shared memory
parallelization; \textbf{OpenMP} The C++ implementation using Intel's OpenMP for
shared memory parallelization; \textbf{SHARK} ExaSHARK version; and
\textbf{GraphLab} Version using GraphLab

The number of latent features ($K$) is equal to 50.

The results show that all parallel implementations of the BPMF scale with the
increasing number of used cores. However there is a clear correlation between
the abstraction level used and the performance obtained. The TBB and OpenMP
versions are the most low-level and obtain highest performance, higher-level
libraries like ExaSHARK and GraphLab focus less on performance and this
gap is clearly visible in the graph. GraphLab, for example, uses TCP sockets
and ehternet instead of MPI and InfiniBand.

%Since the gap is widening when more
%parallel resourses are added, we will no longer compare to GraphLab
%in the distributed versions later.

The TBB version performs better than the OpenMP version because TBB's support
for nested parallelism and because TBB uses a work-stealing scheduler that can
better balance the work.

\subsection{Distributed BPMF} \label{Distributed BPMF}

In this section, the strong scaling of the different versions of distributed BPMF
is studied. We first present results for the ChEMBL dataset on a relatively
small cluster with 12 nodes, comparing the different MPI, GASPI and ExaSHARK
versions, showing the benefit of asynchronous communication even at such small
scales. Then we show that there are large differences between the different
asynchronous versions for larger clusters, and we find the limits of scaling
such a tightly integrated algorithm as BPMF.

\begin{figure*}
    \centering
    \includegraphics[width=0.49\textwidth]{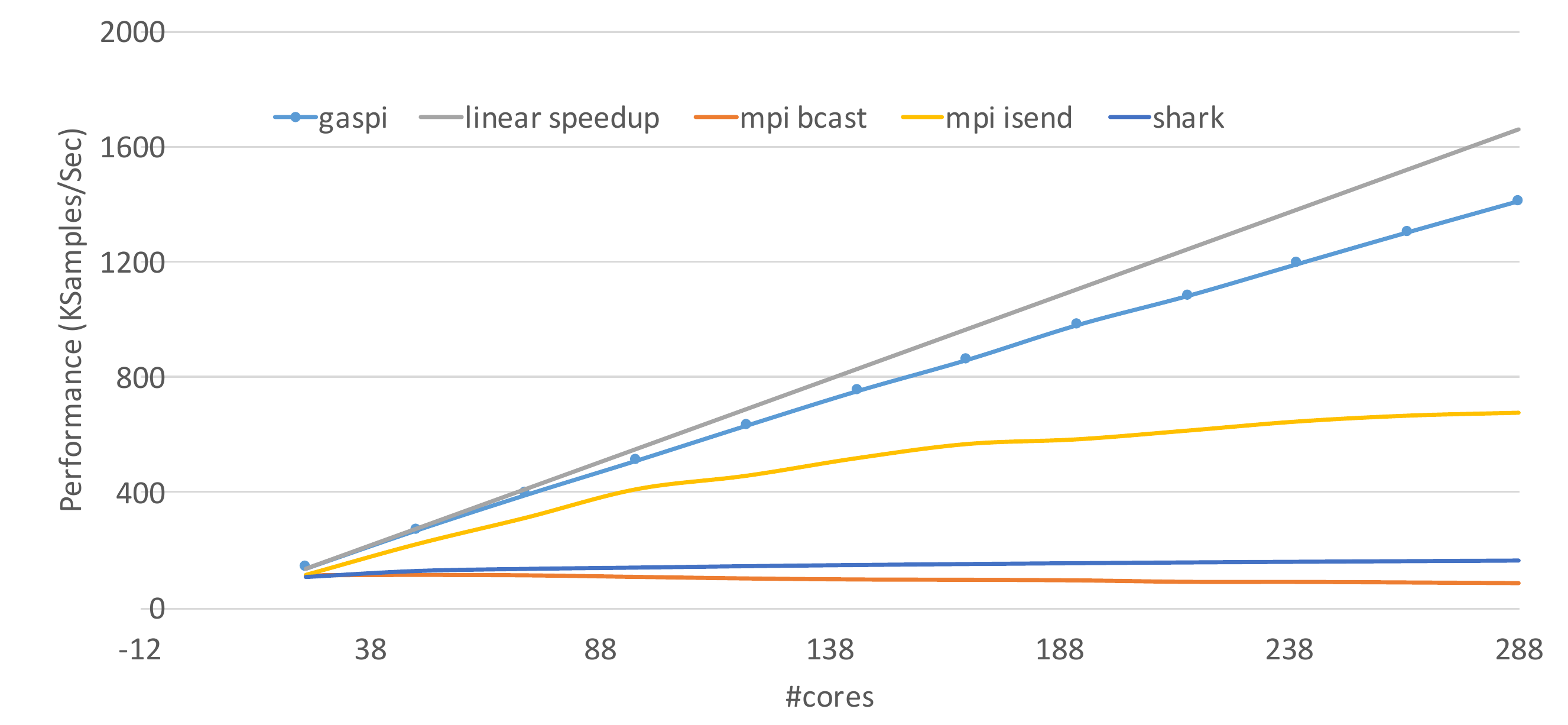}
    \includegraphics[width=0.49\textwidth]{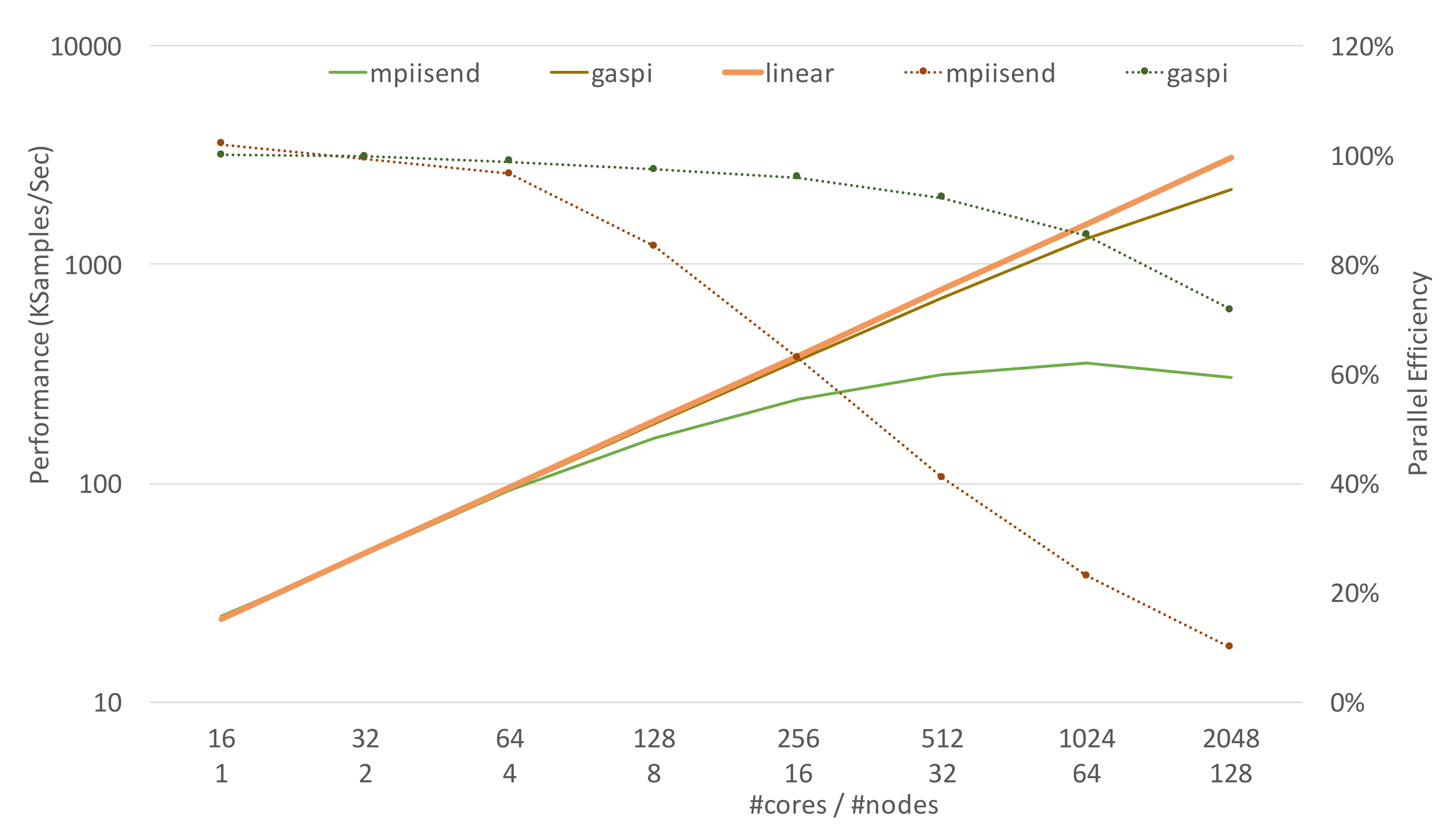}
\caption{Performance of the distributed BPMF on the ChEMBL dataset (left) and MovieLens dataset (right) in number of
updates to $U$ and $V$ per second versus the number of cores used.}
\label{fig:multinode}
\end{figure*}

Figure~\ref{fig:multinode} (left) shows a clear advantage of two asynchronous
communication version being the GASPI version and the MPI version using
\texttt{\small MPI\_Isend} and \texttt{\small MPI\_Irecv}. For these version communication
happens in the background, in parallel with computation, while for the two other
versions, the ExaSHARK version and the version using MPI broadcast
(\texttt{\small MPI\_bcast}) communication is happening after the computation and thus
the performance gained by adding more nodes, is lost again by the time spent
communicating. 

\begin{figure*}
\centering
\includegraphics[width=0.4\textwidth]{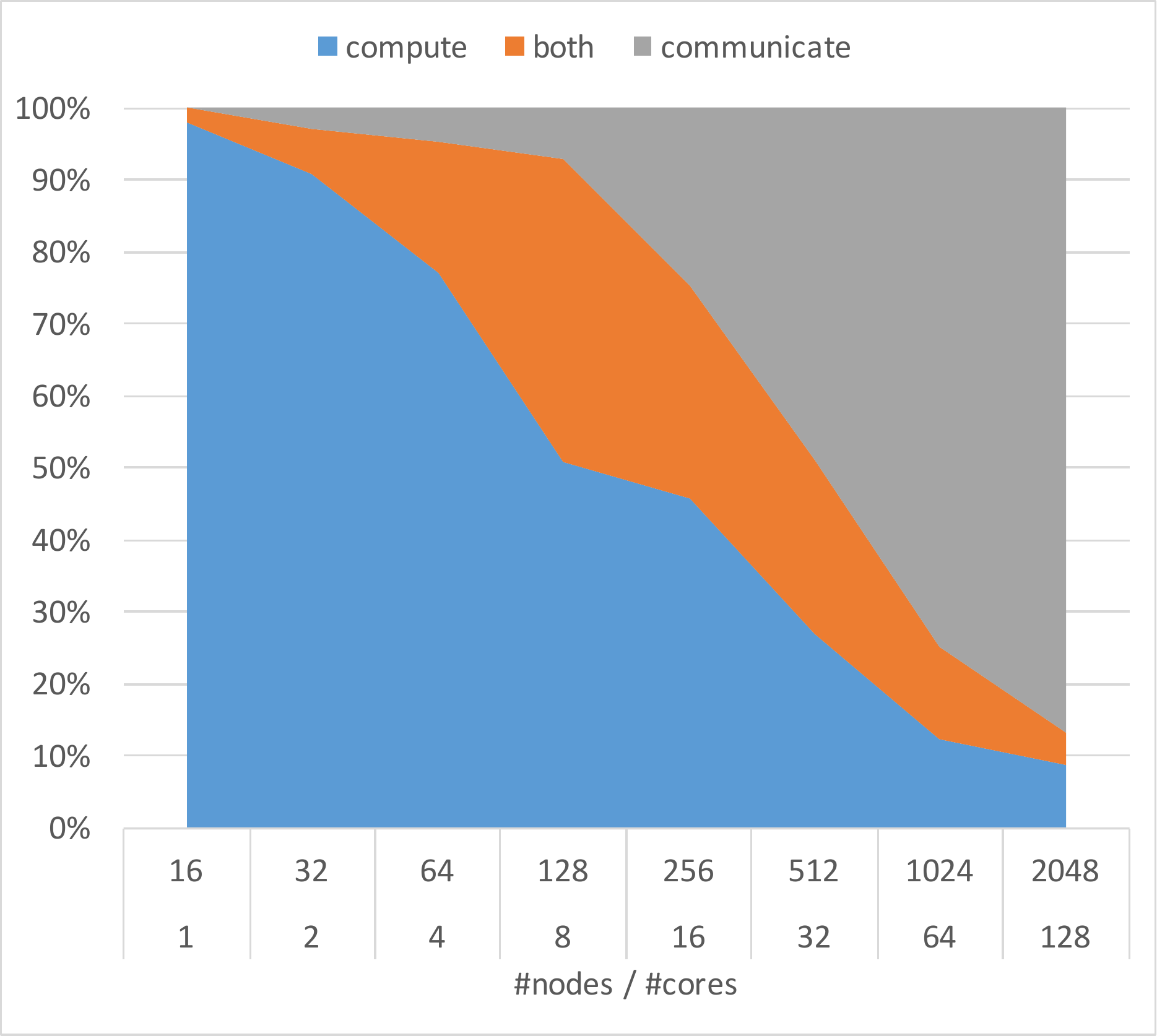}
\includegraphics[width=0.4\textwidth]{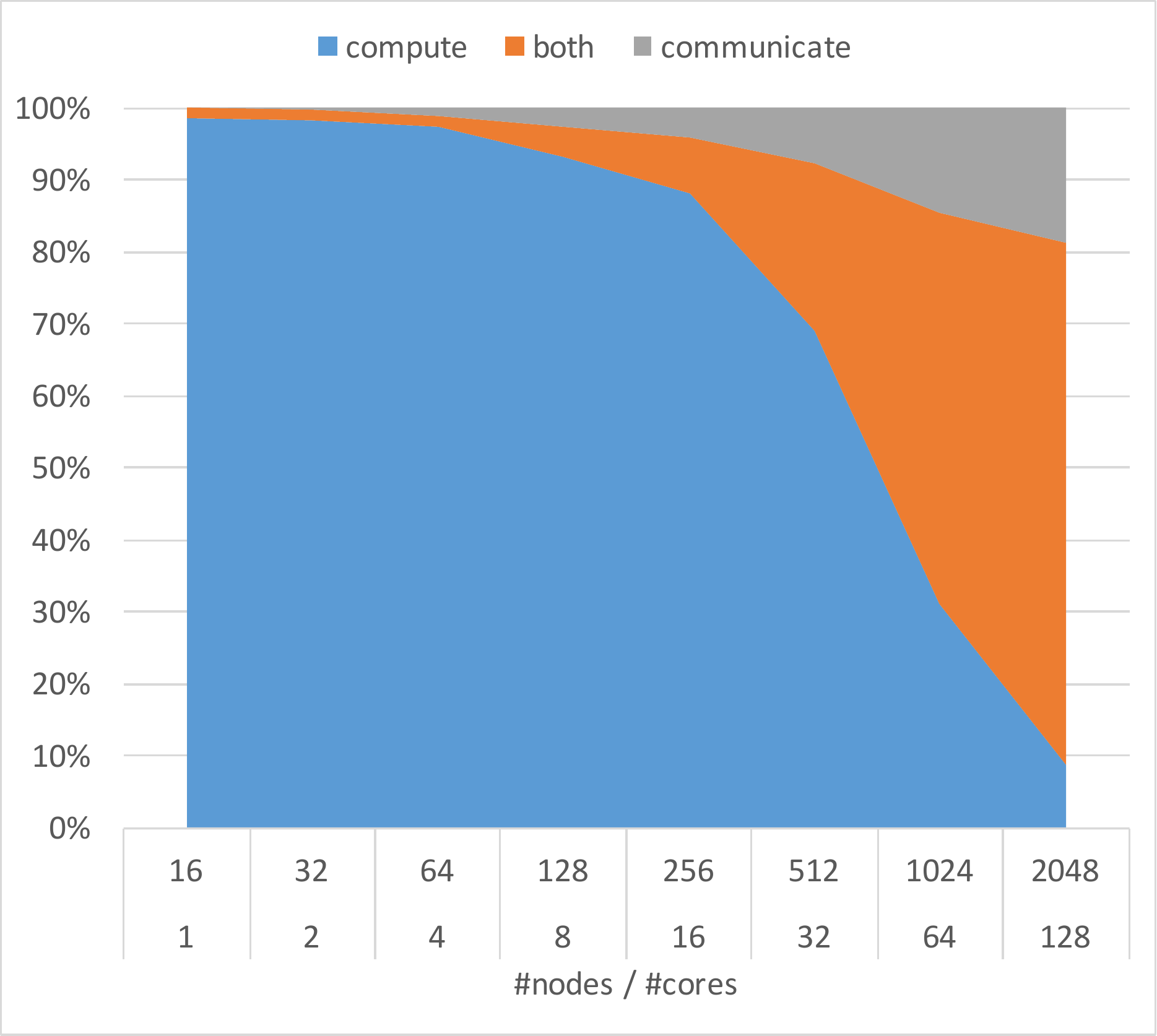}
\caption{Time spent computing, communicating and doing both for the MPI implementation (left) and GASPI implementation (right).}
\label{fig:overlap}
\end{figure*}

Scaling further to 128 nodes, the difference between the asynchronous versions
becomes apparent. Figure~\ref{fig:multinode} (right) shows the GASPI version scales
better than the asynchronous MPI version, achieving more than 70\% parallel
efficiency for 128 nodes compared to 10\% for the MPI version. This is due to
two factors. Firstly the GASPI communication library is much more light-weight
than the MPI version, spending about 2.5x less time than MPI per message sent.
And because of this, secondly the GASPI version allows you to hide 85\% of the
communication time (for 128 nodes), while for the MPI version this is a mere
10\%. The overlap of communication and computation is displayed in
Figure~\ref{fig:overlap}. In this figure \emph{both} means that the network
hardware is sending data (communicating) while the processor is busy doing
computations.  A clear difference between MPI on the left and GASPI on the
right is visible.

As can already be seen from the GASPI results on 128 nodes, we also expect the
performance of the GASPI version to level off. This is due to the general
decrease in the amount of work per node (less items) and increase in the amount
of communication (more nodes). We need changes to the algorithm itself to keep
scaling.

\section{Related Work} \label{sec:related}

% Add recent publications from Google/MIT

Apart from Bayesian Probabilistic Matrix Factorization (BPMF) \cite{BPMF}, the
most popular algorithms for low-row matrix factorization are probably alternating
least-squares (ALS) \cite{parallelALS} and stochastic gradient descent (SGD)
\cite{parallelSGD}.

SGD randomly loops through all observed interactions user-movie, computes the
error of the prediction for each interaction and modifies the model parameters
in the opposite direction of the gradient.  The ALS technique repeatedly keeps
one of the matrices $U$ and $V$ fixed, so that the other one can be optimally
re-computed. ALS then rotates between re-computing the rows of $U$ in one step
and the columns of $V$ in the subsequent step. The advantage of BPMF is that the
predictions are averaged over all the samples from the posterior distribution
and all the model parameters are integrated.

While a growing number of works studied parallel implementations of the SGD
\cite{parallelSGD,parallelSGD2} and ALS \cite{parallelALS}, less research work
dealt with a parallelization of the BPMF \cite{BPMFDesc,parallelMCMC}. Indeed,
computing the posterior inference which time complexity per iteration is cubic
with the respect of the rank of the factor matrix ($\approx$$K$$^3$), may
become very exorbitant when the number of users and movies runs into millions.
SGD, in the other hand, is computationally less expensive even if it needs more
iterations to reach a good enough prediction and its performance is sensitive
to the choice of the learning rate. For ALS, although its time complexity per
iteration, previous related work \cite{parallelALS} showed that it is well
suited for parallelization.

In \cite{parallelMCMC}, a distributed Bayesian matrix factorization algorithm
using stochastic gradient Markov Chain Monte Carlo (MCMC) is proposed. This work
is much more similar to this work than the aforementioned ALS and SGD. In the
paper, the authors extended the Distributed Stochastic Gradient Langevin
Dynamics (DSGLD) for more efficient learning. For the sake of increasing
prediction's accuracy, they use multiple parallel chains in order to collect
samples at a much faster rate and to explore different modes of parameter
space. In this work, the Gibbs sampler is used because it is popular for its
best quality samples even though it is more difficult to parallelize.

From parallel programming prospective, a master slave model is considered in
\cite{parallelMCMC}. The initial matrix $R$ is grid into as many independent
blocks as used workers. At each iteration, the master picks a block using a
block scheduler and sends the corresponding chunk of $U$ and $V$ to the block's
worker. Upon reception, the worker updates these chunks by running DSGLD using
its local block of ratings. Afterwards, the worker sends the chunks back to the
master. Upon reception, this later updates its global copy of the matrices $U$
and $V$. Two levels of parallelism are used by the authors as a way of
compensating the low mixing rate of SGLD: a parallel execution of the same
sampling step (chain) and different samples in parallel. 

In this work, a PGAS approach is used where the computation is totally
decentralized and where the matrices are defined as global arrays. In such a
decentralized model, no global barrier is needed to update the matrices neither
for synchronizing the block distribution scheduling such as in
\cite{parallelMCMC}. No bottleneck is also created when the updates of the
matrices are exchanged.

% The work distribution for the $U$ matrix (the matrix of the users) is
% row-by-row which means that the work sharing is more likely to balanced than a
% block-based distribution. Indeed, the datasets for recommender systems are
% known to be very spare which means that some blocks may have many samples to
% compute while others much less to do. For the distribution of the matrix $V$
% (the matrix of the movies) a work-sharing approach based on user-movie
% interaction is proposed. Even though communicating data is unavoidable for each
% approach, we think that exchanging updates is a decentralized way is more
% scalable than using master-slave fashion.

\section{Conclusion and Future Work}
\label{sec:conclusion}

This work proposed a high-performance distributed implementation of the
Bayesian probabilistic matrix factorization algorithm.  We have shown that load
balancing and low-overhead asynchronous communication are essential to achieve
good parallel efficiency, clearly outperforming more common synchronous
approaches like GraphLab.  The achieved speed-up allowed us to speed up machine
learning for drug discovery on an industrial dataset from 15 days for the
initial Julia-based version to 5 minutes using the distributed version with TBB
and GASPI.

Future work includes extending the framework to support more matrix
factorization methods such as Group Factor Analysis \cite{GFA} or Macau
\cite{simm:macau}, but also a look at more scalable MF \emph{algorithms}.

\section*{Acknowledgments}
This work is partly funded by the European project ExCAPE with reference
671555.  We acknowledge IT4I for providing access to the Anselm and Salomon
systems.

\end{document}